\documentclass[aps,prb,twocolumn,superscriptaddress]{revtex4-2}
\usepackage{graphicx}
\usepackage{latexsym}
\usepackage{changes}
\usepackage{amssymb}
\usepackage{amsmath}
\usepackage{amsfonts}
\usepackage{upgreek}
\usepackage{subfigure}
\usepackage{bm}
\usepackage{bbold}
\usepackage[unicode=true,
 bookmarks=false,
 breaklinks=false,pdfborder={0 0 1},backref=false,colorlinks=true]
 {hyperref}
\hypersetup{
 linkcolor=[rgb]{0,0,1},citecolor=[rgb]{0,0,1},urlcolor=[rgb]{0,0,1}}
\usepackage{multirow}
\usepackage{color}
\usepackage{comment}
\usepackage{xcolor}
\usepackage{soul}
\usepackage{tikz}

\usepackage{braket}

\usepackage{tabularx}
\newcolumntype{L}{>{$}l<{$}} 

\begin{document}

\title{Bulk topological signatures of a quasicrystal}

\author{Gautam Rai}\email{gautamra@usc.edu}
\author{Henning Schl\"{o}mer} \altaffiliation[Also at ]{RWTH Aachen University}
\author{Chris Matsumura}%
\author{Stephan Haas} \altaffiliation[Also at ]{Jacobs University Bremen}
\affiliation{%
 Department of Physics and Astronomy, University of Southern California
}%

\author{Anuradha Jagannathan}
\affiliation{
Laboratoire de Physique des Solides, Universit\'{e} Paris-Saclay
}%

\date{\today}

\begin{abstract}
We show how measuring real space properties such as the charge density in a quasiperiodic system can be used to gain insight into their topological properties. In particular, for the Fibonacci chain, we show that the total onsite charge oscillates when plotted in the appropriate coordinates, and the number of oscillations is given by the  topological label of the gap in which the Fermi level lies. We show that these oscillations have two distinct interpretations, obtained by extrapolating results from the two extreme limits of the Fibonacci chain---the valence bond picture in the strong modulation limit, and perturbation around the periodic chain in the weak modulation limit. This effect is found to remain robust at moderate interactions, as well as in the presence of disorder. We conclude that experimental measurement of the real space charge distribution can yield information on topological properties in a straightforward way.
\end{abstract}

\maketitle

\section{Introduction} Quasicrystals have recently been shown to exhibit rich non-trivial topological phases \cite{krausQuasiperiodicityTopologyTranscend2016, flickerQuasiperiodicity2DTopology2015}. We consider in particular the pure hopping Fibonacci Hamiltonian, which is known to have gaps everywhere in the spectrum. One can assign a topological label to each of them by virtue of the gap-labeling theorem \cite{bellissardGapLabellingTheorems1992,jagannathanFibonacciQuasicrystalCase2020}. These gap labels map directly to the Chern numbers of an appropriately extended model \cite{krausTopologicalEquivalenceFibonacci2012}. In this paper, we show how these labels can be determined from relatively simple experimental measurement of the electronic charge distribution along such a Fibonacci chain.
Our proposed measurement is distinct from previous studies, which characterize the topology via the the winding of edge modes as a function of an external tuning parameter. For example, in a recent polaritonic crystal experiment \cite{babouxMeasuringTopologicalInvariants2017}, the external parameter is the phason angle $\phi$, which progressively transforms the Fibonacci chain by a series of bond flips. It was shown that the energy of an edge state in a gap with label $q$ crosses the gap $|q|$ times as a function of $\phi$. Every gap crossing is accompanied by the corresponding edge state swapping its localization. This idea has been used to build an adiabatic pump \cite{verbinTopologicalPumpingPhotonic2015}, whereby charge was pumped from one edge of a photonic crystal to the other by smoothly modulating between Fibonacci potentials with different values of $\phi$. These (and related \cite{raiProximityEffectSuperconductorquasicrystal2019,dareauRevealingTopologyQuasicrystals2017}) studies probe topological quantities by studying the behavior of edge states under variation of a control parameter, namely the phason angle. Given the bulk-edge correspondence principle, the question arises if and how the topology manifests itself in bulk quantities. In this letter, we show this is indeed the case, and furthermore that a single realization of the quasicrystal suffices: it is not necessary to tune an external parameter such as the phason angle.

The key idea is to plot real space quantities such as the charge density, $n_i = \braket{c^{\dagger}_i c_i}$, as a function of their conumbering index \cite{sireExcitationSpectrumExtended1990}. In this ordering, the oscillations of the charge density indicate the label of the gap in which the Fermi level lies. We show that this property is shared by other quantities such as the entanglement entropy, $S(\ell)$, and that it is robust to disorder and interactions. 

Our result is immediately applicable to experimental setups that allow doping-controlled measurements, such as STM manipulation of confined electron surface states \cite{kempkesDesignCharacterizationElectrons2019, slotExperimentalRealizationCharacterization2017} and photoemission spectra of polaritonic \cite{babouxMeasuringTopologicalInvariants2017, taneseFractalEnergySpectrum2014} and photonic \cite{verbinTopologicalPumpingPhotonic2015} crystals. 

%
%
\section{A brief introduction to the Fibonacci chain}
\begin{figure}
    \centering
    \includegraphics[width = \columnwidth, trim = 0 40pt 0 40pt, clip]{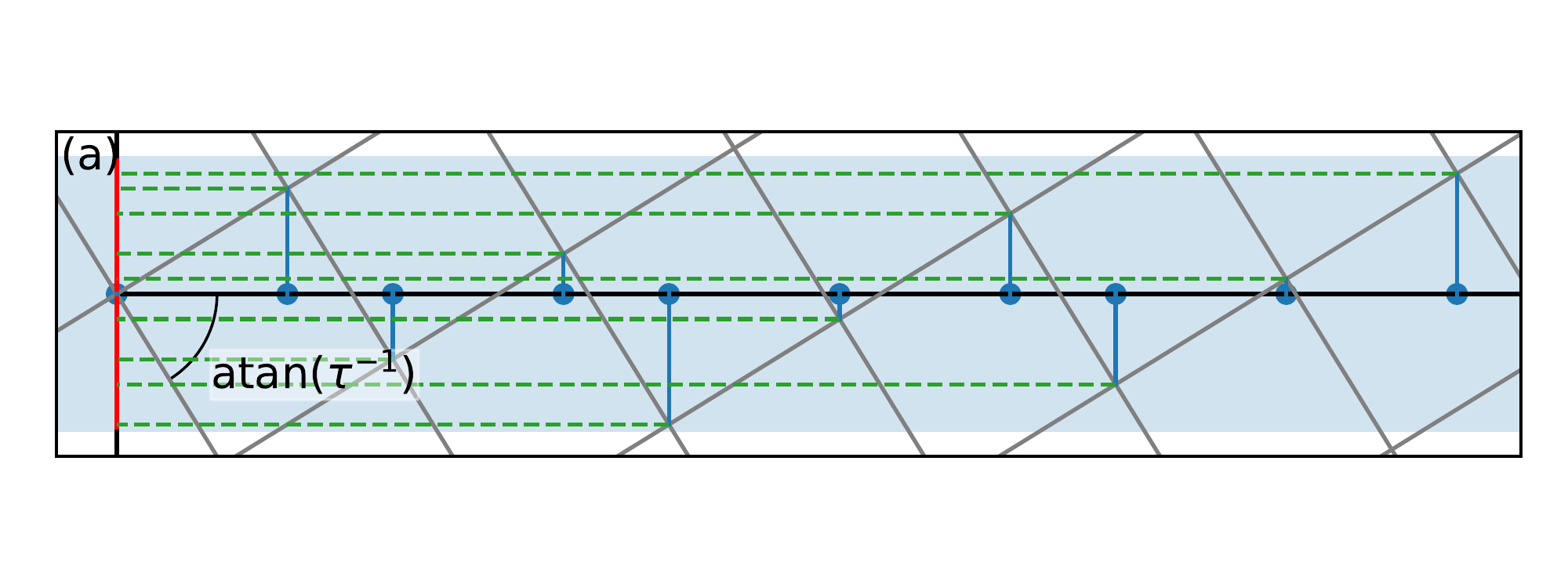}
    \includegraphics[width = \columnwidth]{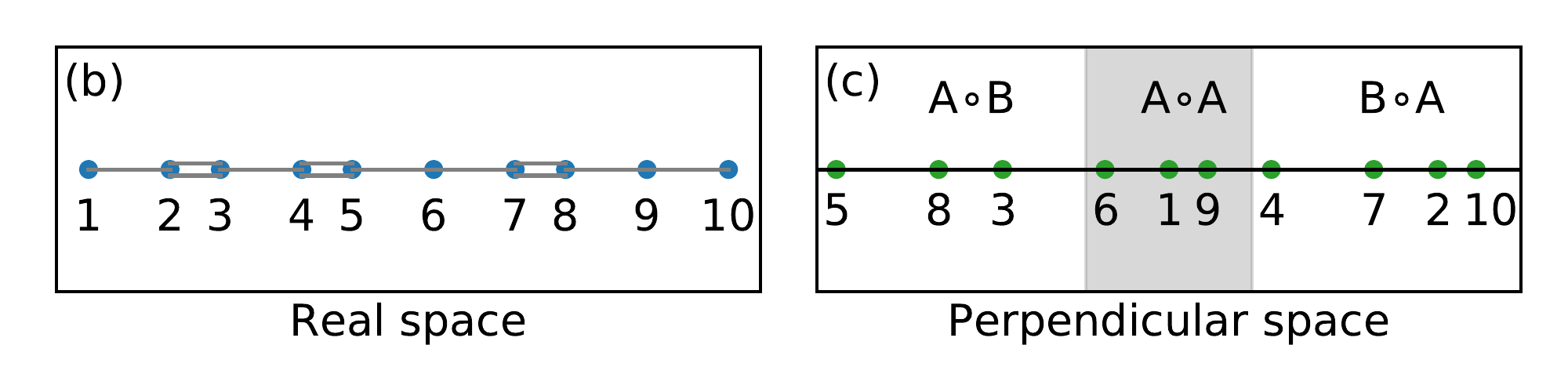}
    \caption{(a) Cut-and-project method to obtain the Fibonacci chain: points in a regular 2D grid are projected onto the \emph{physical line} of slope $\frac{1+\sqrt{5}}{2}$. The projections (solid lines) represent sites of the quasicrystalline lattice. The projections onto a plane perpendicular to the physical line (dashed lines) define the co-number of each site. (b) The sites corresponding to (a) in real space. (c) The reordering of the sites in (b) according to their projection in perpendicular space: also known as co-numbering.}
    \label{fig:cutandproject}
\end{figure}
The Fibonacci chain is a 1D quasicrystal obtained by projecting points from the 2D square grid onto the \emph{physical line} of slope $\tau = \frac{1+\sqrt{5}}{2}$ (see Fig. \ref{fig:cutandproject}).
The simplest physical model corresponding to the Fibonacci chain is the Fibonacci hopping model,
\begin{align}\label{eq:Hamiltonian}
    \hat{H} = -\mu\sum_i c^\dagger_i c_{i} - \sum_i t_i c^\dagger_i c_{i+1} + h.c.,
\end{align}
where the nearest neighbor hopping integrals $t_i$ take two values $t_A$ or $t_B$ according to the Fibonacci sequence. $\mu$ is the chemical potential which is set to be constant throughout the chain.
\begin{figure*}
    \includegraphics[width = 0.4\textwidth]{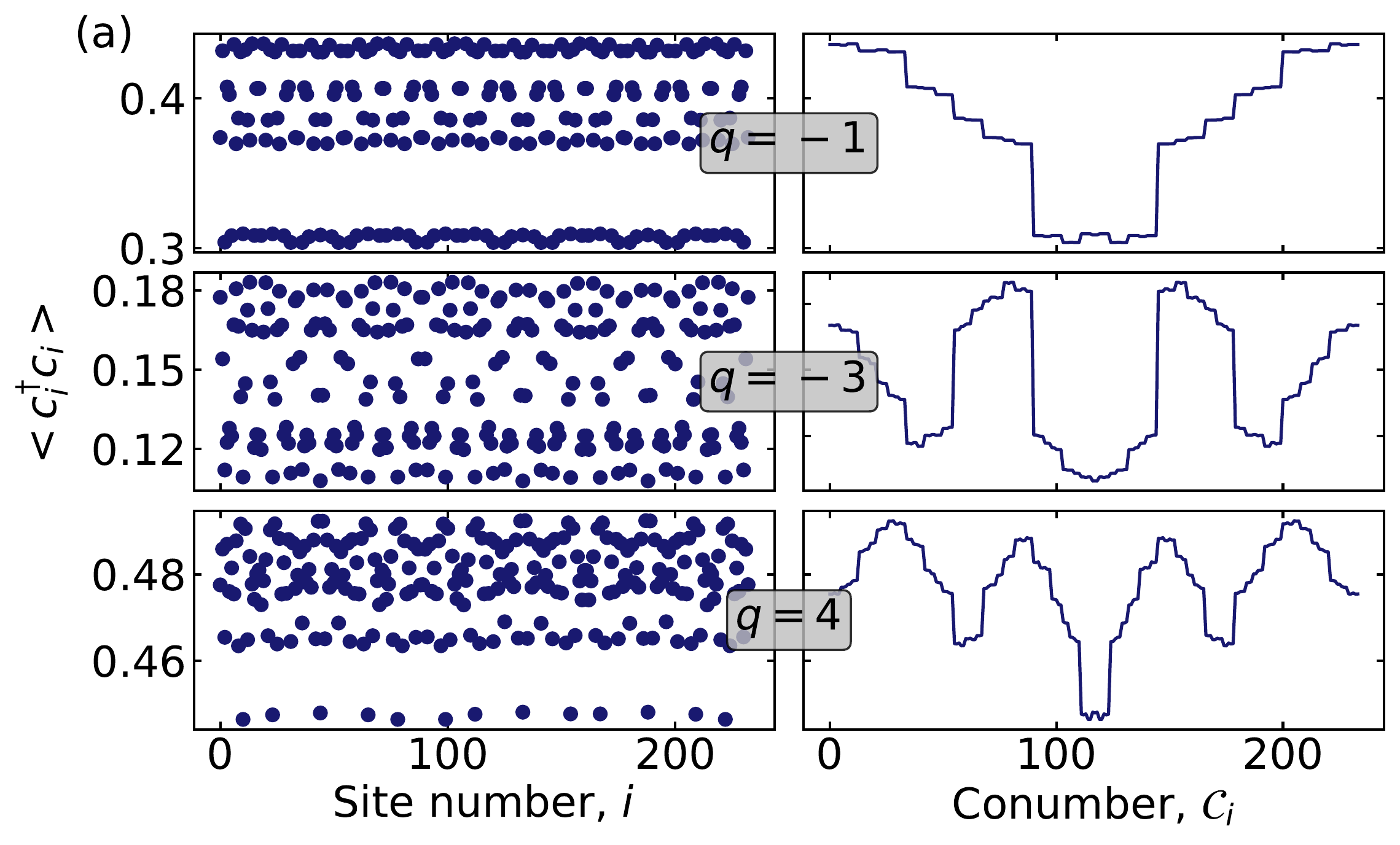}\includegraphics[width = 0.26\textwidth]{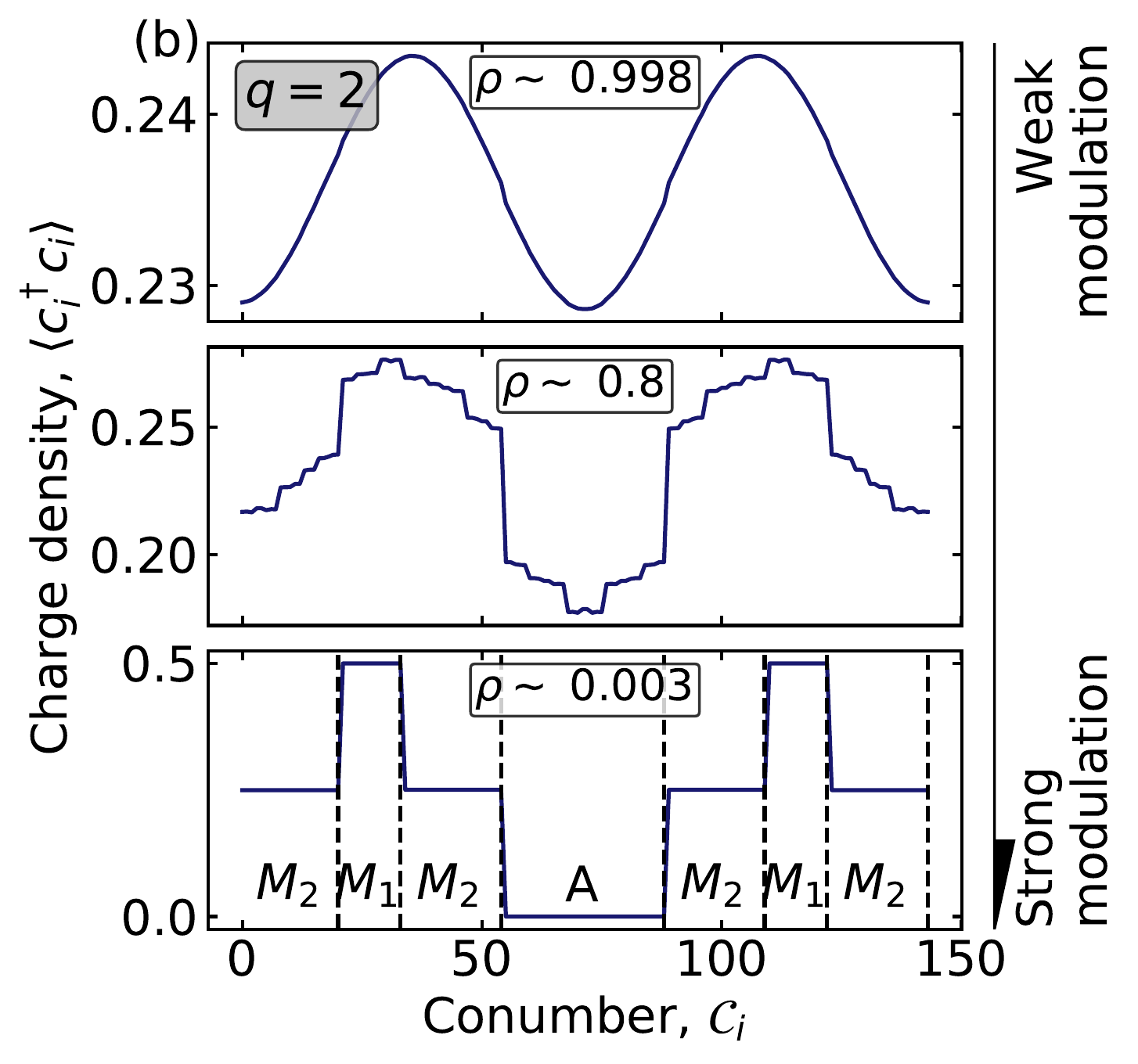}\includegraphics[width = 0.34\textwidth]{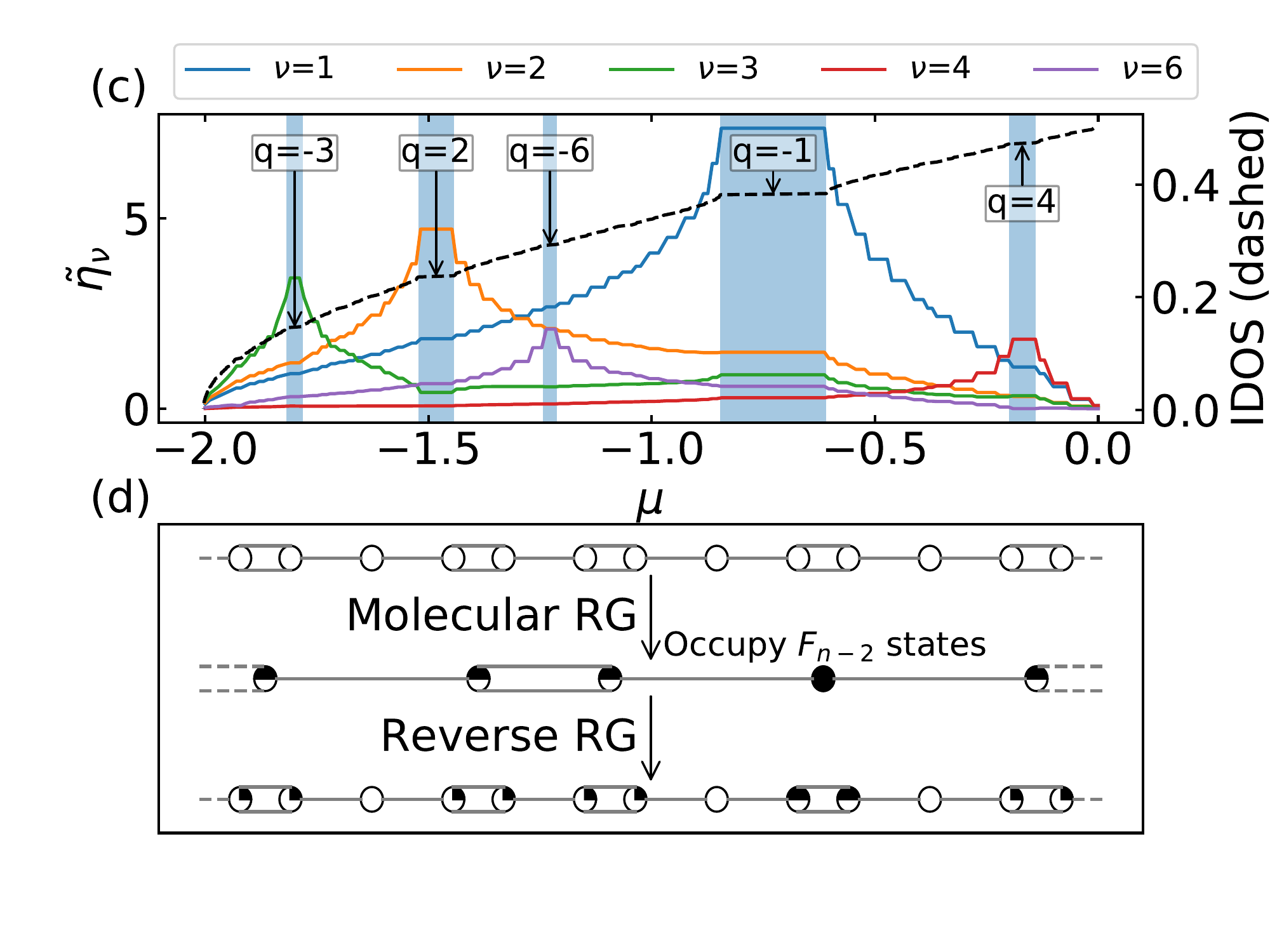}
    \caption{(a) The charge density, $\braket{c_i^{\dagger} c_i}$, in real space (left) and perpendicular space (right), with the Fermi level tuned to the $q = -1, -3$, and $4$ gaps. (b) The charge density in the $q = 2$ gap for three different values of $\rho$ representing strong, intermediate, and weak modulation. (c) The first few Fourier components of the perpendicular space charge density are plotted as a function of the chemical potential, $\mu$. Also shown is the integrated density of states (dashed line) at energy $\mu$, with the five largest gaps in the lower half of the spectrum marked with their $q$-labels. (d) Example of RG analysis in the strong modulation limit for a $13$-site Fibonacci chain  tuned to the $q=2$ gap.}
    \label{fig:Charge-density}
\end{figure*}

The spectrum of the Fibonacci hopping model in \eqref{eq:Hamiltonian} is a Cantor set. There are an infinite number of gaps, and each gap can be labelled by a unique integer $q$. According to the gap-labelling theorem \cite{jagannathanFibonacciQuasicrystalCase2020}, the number of states below a gap with label $q$ is given by the integrated density of states (IDOS),
\begin{align}
    IDOS(\epsilon)\Big|_{\epsilon\in q\text{-gap}} = q\tau^{-1} \bmod 1 .\label{eq:gap-labelling}
\end{align}
In this paper, we perform  calculations on approximants of the Fibonacci chain. An approximant is a finite structure obtained as the supercell of the periodic chain obtained by the cut-and-project method if the golden ratio $\tau$ is replaced by a convergent $\tau_n = F_n/F_{n-1}$ where $F_n$ is a Fibonacci number. $t_A$ and $t_B$ are chosen such that the average hopping $t=1$ \footnote{We pick $t_A$ and $t_B$ such that the average hopping $t = \frac{F_{n-1}t_A + F_{n-2}t_B}{F_n} = 1$. }. This ensures that the bandwidth remains constant. All energies are given in units of $t$. A modified version of the gap-labelling theorem applies to the approximants,
\begin{align}
    IDOS(\epsilon)\Big|_{\epsilon\in q\text{-gap}} = q\frac{F_{n-1}}{F_n} \bmod 1,\label{eq:gap-labelling-finite}
\end{align}
where $q$ belongs to a finite subset of the integers depending on the size of the approximant.

The 1D Hamiltonian \eqref{eq:Hamiltonian} admits a natural extension to a 2D Hamiltonian of electrons hopping on a rectangular lattice in the presence of a magnetic field perpendicular to the plane \cite{krausTopologicalEquivalenceFibonacci2012}. The $q$-labels of the Fibonacci chain are directly related to the Chern numbers of this \emph{ancestor} 2D Hamiltonian.

The co-numbering scheme is used to obtain an alternate ordering of the sites in the Fibonacci chain \cite{sireExcitationSpectrumExtended1990}. We simply arrange the sites according to the projection of the corresponding 2D lattice points onto the \emph{perpendicular space}---a line perpendicular to the physical line (see Fig. \ref{fig:cutandproject}). By this procedure, sites that have more similar local environments in real space are placed closer together in  perpendicular space. For example, sites in the central cluster in perpendicular space have an $A$ bond on each side, whereas sites in the left(right) cluster have an $A(B)$ bond to the right and a $B(A)$ bond to the left. We can further distinguish sites within each cluster by comparing successively larger local neighborhoods of the sites. By numbering all the projections according to their positions in perpendicular space from left to right, we can define the co-number, $j$, of the $i$th site of a given Fibonacci approximant: $\mathcal{C}:\mathbb{Z}\to\mathbb{Z}$. 

\section{Charge density oscillations in perpendicular space} The main result of this paper is that the topological invariants of the Fibonacci chain, $q$, can be observed directly by measuring the charge density of the the Fibonacci chain. The charge density, $n_i$, is inhomogeneous and is given by
\begin{equation}
n_i = \braket{c^\dagger_i c_i} = \sum_{k} c^\dagger_i c_i f(\epsilon_k-\epsilon_f, T),
\end{equation}
where $i \in \{0,1,2,\ldots F_{n}-1\}$ labels each site according to its position along the chain, $f(\epsilon, T)$ is the Fermi distribution function at temperature $T$, and $\epsilon_k$ are the eigenvalues of the Hamiltonian. The key idea is to map the charge density to perpendicular space, $\eta_j = n_{\mathcal{C}^{-1}(j)}$. This corresponds to reshuffling the sites such that they are arranged according to their positions in perpendicular space.

The tight-binding calculations were performed assuming periodic boundary conditions \footnote{We used the Python package Kwant \cite{grothKwantSoftwarePackage2014} to set up the Hamiltonian matrix}. Fig.~\ref{fig:Charge-density}~(a) \, shows the charge density at zero temperature in real space as well as in perpendicular space for three choices of $\mu$, corresponding to the Fermi level lying in the $q = -1, -3,\text{ and }4$ gaps. The charge density in perpendicular space reveals the underlying topological structure. $\eta_j$ winds $q$ times when the Fermi level is tuned to the gap with label $\pm q$. This suggests a direct and straightforward way to determine the topological invariants of the Fibonacci chain via a doping controlled measurement of the real space charge density for a single realization. 

We investigate this further by computing the Fourier spectrum of the charge density in perpendicular space,
\begin{align}
    \tilde{\eta}_\nu = \sum_{j = 0}^{F_n - 1} e^{ 2\pi \texttt{i}\frac{j \nu }{F_n}} \eta_j.
\end{align}
In Fig. \ref{fig:Charge-density} (b), the $\nu$th Fourier component, $\tilde{\eta}_\nu$, is plotted as a function of the chemical potential, for $\nu$ corresponding to the $q$-labels of the five largest gaps in the spectrum of the Fibonacci chain $(q = -1, 2, -3, 4, -6)$. The same plot also shows the integrated density of states of the Fibonacci chain, corresponding to the energy $\mu$. The five largest gaps are marked with a shaded background. The magnitude of the $\nu$th Fourier component, $\tilde{\eta}_\nu$, is largest precisely when the Fermi level is tuned to aa gap with  $|q| = \nu$. The effect is particularly prominent for small $q$ (large gaps), where the $q$th component dominates over all the other frequencies. 


We can gain additional insight into these charge density oscillations by considering two opposite limits---the strong modulation limit ($\rho = t_A/t_B\to0$) and the weak modulation limit ($\rho\to1$). The topology of the spectrum is unchanged for the Fibonacci chain as long as $\rho$ is finite and non-zero. All the gaps remain open when modulating from one extreme to the other. The intuition that accompanies each limit is different, but the fundamental observation is the same---the charge density in a $q$-gap oscillates $q$ times in perpendicular space (see Fig.~\ref{fig:Charge-density}~(b)).
\subsection{The weak modulation limit}
In the weak modulation regime, we follow the perturbative analysis of Sire and Mosseri \cite{sireSpectrum1DQuasicrystals1989}, writing the Hamiltonian as a sum of two terms,
\begin{align}
    \hat{H} =& \underbrace{- \sum_{i=1}^{F_n} t_A c^\dagger_i c_{i+1} + hc}_{\hat{H}_0}
    \underbrace{- \sum_{i=1}^{F_n} (t_i - t_A) c^\dagger_i c_{i+1} + hc}_{\hat{H}_w}
\end{align}
$\hat{H}_0$ is the Hamiltonian of a periodic chain with hopping $t_A$, and $\hat{H}_w$ introduces the Fibonacci modulation. We use periodic boundary conditions, so the index $i$ (and $j$ in the following) are understood to wrap around modulo $F_n$. In perpendicular space, these terms take a particularly simple form \footnote{$\hat{\mathcal{H}} = \hat{\mathcal{H}}_0 + \hat{\mathcal{H}}_w$ itself has a simpler form in perpendicular space. It is a symmetric Toeplitz matrix with with 4 filled diagonals (other than the chemical in the main diagonal)},
\begin{align}
    \hat{\mathcal{H}}_0 &= -\sum_{j=1}^{F_n} t_A c_j^\dagger c_{j+F_{n-2}} + t_A c_j^\dagger c_{j+ F_{n-1}} + h.c. \label{eq:H0_perp}\\
    \hat{\mathcal{H}}_w &= -\sum_{j=1}^{F_{n-2}} 2w c_j^\dagger c_{j+F_{n-1}} + h.c. \label{Hw_perp}
\end{align}
where we have used $2w = t_A-t_B$. $\hat{\mathcal{H}}_w$ connects each of the leftmost $F_{n-2}$ sites to the site $F_{n-1}$ places to its right. 

\begin{figure}
    \centering
    \includegraphics[width = \columnwidth]{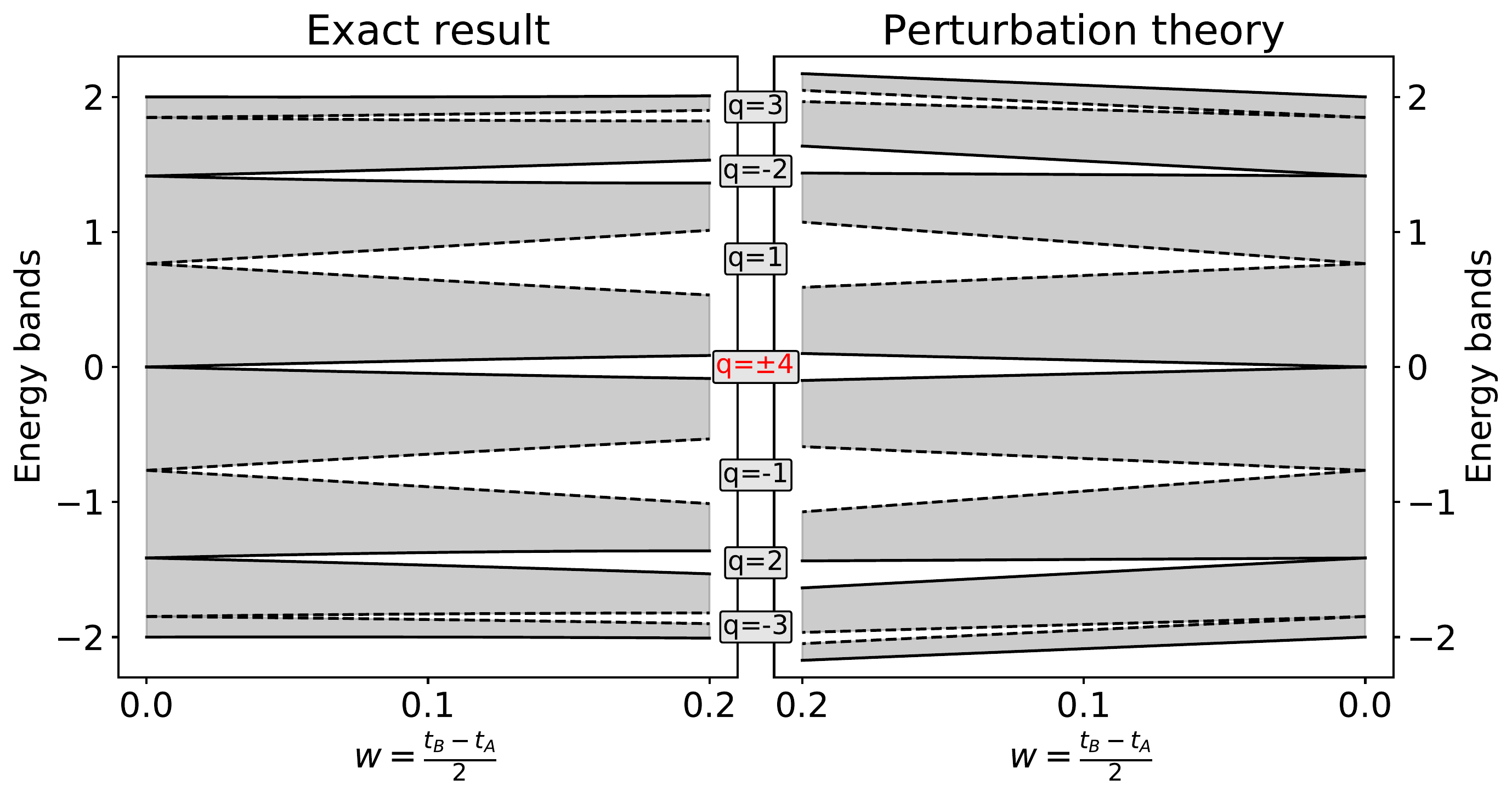}
    \caption{The opening of gaps with increasing modulation strength in the band structure of the $F_n = 8$ approximant: a comparison of first order perturbation theory with the exact result. The solid lines are the eigenvalues at the center of the Brillouin zone and the dashed lines are the eigenvalues at the edges of the Brillouin zone. The shaded regions are the bands.}
    \label{fig:exactvsperturbation}
\end{figure}

$\mathcal{H}_0$ can labelled by the integers $\mu = 0,1,\ldots,\lfloor F_n/2\rfloor$ and its eigenvalues are given by $\mathcal{E}_\mu = -2t_A\cos\left(\frac{2\pi F_{n-1}}{F_{n}}\mu\right)$. They are (almost all) doubly degenerate and the corresponding eigenvectors are given by $v_\mu(j) = \frac{1}{\sqrt{F_n}}\exp\left(\pm\frac{2\pi i }{F_{n}}\mu j\right)$. Notice that the eigenvectors of the periodic chain, which are plane waves in real space, are also plane waves in perpendicular space. The wavenumber of the plane wave in real space $m$ is related to the wavenumber in perpendicular space $\mu$ by equating $\cos\left(\frac{2\pi m}{F_{n}}\right) = \cos\left(\frac{2\pi F_{n-1}\mu}{F_{n}}\right)$.
\begin{align}\label{eq:mcongruentmu}
    m \equiv \mu F_{n-1} \pmod{F_n}
\end{align}
In \cite{sireSpectrum1DQuasicrystals1989}, the authors use first order perturbation theory to show that the perturbation  $\hat{\mathcal{H}}_w$ immediately opens the gaps of the Fibonacci chain. The opening of the gap at energy $\mathcal{E}_\mu$ is associated with the splitting of the degenerate plane wave states. To first order, the perturbed energies and states around a gap that has an even number of states below it are given by (see appendix for derivation)
\begin{align}
    \mathcal{E}_\mu^{\pm} =& -\left( 2t_{A} + 4w\frac{F_{n-2}}{F_{n}}\right) \cos\left(\frac{2\pi F_{n-1}}{F_{n}}\mu\right) \nonumber\\
    &\mp \frac{4w}{F_{n}} \frac{\sin\left( 2\pi \frac{F_{n-2}}{F_{n}}\mu\right)}{\sin\left( \frac{2\pi }{F_{n}}\mu\right)},\label{eq:pert_eps}\\
    v_\mu^+(j) =& \sqrt{\frac{2}{F_n}}\cos\left(\frac{2\pi \mu j}{F_n}\right),
    v_\mu^-(j) = \sqrt{\frac{2}{F_n}}\sin\left(\frac{2\pi \mu j}{F_n}\right). \label{eq:pert_vec}
\end{align}
{If there are an odd number of states below a gap, the same expressions apply except $\mu\to\mu+\frac{1}{2}$.} The opening of the gaps as the perturbation is turned on is shown in Fig.~\ref{fig:exactvsperturbation}.

Comparing \eqref{eq:mcongruentmu} with the gap labelling theorem \eqref{eq:gap-labelling-finite}, we obtain that the slightly perturbed plane waves around the gaps with an even (odd) number of states below them have the wave number $\mu = \frac{cF_n + q}{2}$ $(\mu = \frac{cF_n + q -1}{2})$ where $q$ is the gap label and $c$ is some integer. If the Fermi level is tuned to a gap $q$, then only one of a pair of states with this wave number is occupied. The contribution of this state to the charge density $\eta_j^q$ is given by the squared amplitude of one of $v_\mu^\pm(j)$:
\begin{align}
    \eta_j^{q} \propto 
    \frac{1}{2}\left(1\pm\cos\left(\frac{qj}{F_n}\right)\right)
\end{align}
All other occupied states appear in pairs, so their contribution to the charge density is constant.  This is precisely the behavior seen in the top panel of Fig.~\ref{fig:Charge-density}(b). The oscillations are sinusoidal with $q$ periods as the conumber $j$ varies in the interval $[0,F_n-1]$. They arise from the fact that the unperturbed perpendicular space eigenstates at the energy where the gap opens are plane waves with exactly the wave vector needed to obtain $q$ oscillations.
\subsection{The strong modulation limit}
When $\rho = 0$, a Fibonacci approximant of length $F_n$ breaks into $F_{n-3}$ atom sites (with an $A$ bond on both sides) and $2F_{n-2}$ molecule sites (with a $B$ bond on either side). There are three kinds of electronic states available: $F_{n-2}$ (anti-)bonding molecular states with energy $\pm t_B$, and $F_{n-3}$ atomic states with zero energy. When $\rho$ is slowly turned on, the self-similar eigenvalue spectrum can be built by splitting each cluster of states into three subclusters recursively \cite{kaluginElectronSpectrumOnedimensional1986}. The charge density in the strong modulation limit can be recovered by considering the real space structure of all the states below a certain gap. We show how this can be done by carefully applying Niu and Nori's renormalization group method \cite{niuRenormalizationGroupStudyOneDimensional1986}.

We illustrate this with the example of a $13$-site Fibonacci chain tuned to the $q=2$ gap in Fig. \ref{fig:Charge-density} (d). By \eqref{eq:gap-labelling-finite}, this filling corresponds to $3$ electrons in the system. To place these electrons unambiguously, we first perform one step of the molecular RG detailed in \cite{niuRenormalizationGroupStudyOneDimensional1986} resulting in a Fibonacci chain with $4$ molecule sites and $1$ atom site. We fill the $3$ electrons into the lowest energy states of this reduced system, which are the $2$ bonding states followed by the single atom state. Finally, we reverse the decimation step, distributing the electron weight equally between the two ancestor molecule sites in the original chain. This corresponds exactly to the picture in the bottom panel Fig. \ref{fig:Charge-density} (b). We find a charge density of $0.5$ in  regions of perpendicular space populated by molecule sites in molecules with atom sites to both sides (labelled by $M_1$). These are precisely the molecule sites that become atom sites after one decimation step. Other molecule sites ($M_2$) have a charge density of $0.25$ and atom sites ($A$) have a charge density of $0$.

We can formulate a general recipe for obtaining the charge density profile in this way. When $q$ is a Fibonacci number $F_m$, the number of electrons in an $F_n$-site Fibonacci approximant takes a simple form: $F_{n-m-1}$. If $m$ is odd \footnote{We use the convention $F_1 = 1$ and $F_2 = 2$}, first use the molecular RG step $\frac{m-1}{2}$ times to produce a chain with $F_{n-m+1}$ sites and place an electron in each of the $F_{n-m-1}$ bonding states. Then reverse each of the RG steps while distributing any electron weight evenly among among ancestor sites. In the case of even $m$, apply the molecular RG step $\frac{m}{2}$ times, then place one electron in all the bonding states as well as in all of the atom states. Reverse the deflation steps in the same way as before. This procedure can be  generalized to arbitrary $q$ by applying Zeckendorf's theorem \cite{brownZeckendorfTheoremAppucations1964}. When an $F_n$-site Fibonacci chain is tuned to a gap with an arbitrary label $q$, the number of electrons in the system, $I_q$, may not be a Fibonacci number. However, Zeckendorf's theorem guarantees that we can uniquely express any positive integer as a sum of distinct Fibonacci numbers with the constraint that the sum does not include any two consecutive Fibonacci numbers. We write $I_q = F_{m_1} + F_{m_2} + \ldots + F_{m_M}$ with the constraint $|m_\alpha - m_\beta|\geq 2$ and $m_1>m_2>m_3 \ldots >m_M$. We can now split the problem of filling $I_q$ electrons in the original Fibonacci chain to $M$ independent problems of filling $F_{m_\alpha} (\alpha =1,2,\ldots,M)$ electrons in Fibonacci chains of different lengths.

This is done recursively. In the first step, we follow the procedure outlined above to fill $F_{m_1}$ electrons in the original Fibonacci chain but stop before the final step of reversing the molecular RG. If $m_1$ was odd, we now have an $F_{n-m_1+1}$-site Fibonacci chain with all the molecular bonding states filled. Now, the question of how the remaining electrons are distributed among the remaining atom sites is an independent problem. This can be formalized by performing one decimation step of the atomic RG \cite{niuRenormalizationGroupStudyOneDimensional1986} which keeps only the atom sites. The new nearest neighbors are connected with a renormalized strong bond if there used to be one molecule between them, and with a renormalized weak bond if there used to be two molecules between them. Now, the problem has reduced to filling $F_{m_2} + F_{m_3} + \ldots + F_{m_M}$ electrons in a $F_{n-m_1-2}$-site Fibonacci. Alternatively, if $m_1$ was even we are left with an $F_{n-m_1+1}$-site Fibonacci chain with all the molecular bonding states and atom states filled. As before, the problem of distributing the remaining electron weight of $F_{m_2} + F_{m_3} + \ldots F_{m_M}$ electrons among the anti-bonding states is an independent problem which can be formalized by performing one molecular RG step. By performing this procedure $M$ times and then reversing every RG step, taking care to distribute any electron weight in a deflated chain evenly among the ancestor states we recover the charge density profile in the original chain in the strong modulation limit.

This procedure demonstrates that in the strong modulation limit, the charge density at a given site depends only on a small local neighborhood around it. The size of the relevant local neighborhood for a given gap depends on the number of RG steps needed to fill in the electrons up to that gap, which is larger for small gaps with large $q$-labels. Distinguishing large local neighborhoods requires resolving smaller regions in perpendicular space, which results in faster oscillations in perpendicular space. 

The charge density in the intermediate regime interpolates between the two regimes and exhibits characteristics of both pictures, as shown in  Fig~\ref{fig:Charge-density}~(a).
\begin{figure}
    \centering
    \includegraphics[width = \columnwidth]{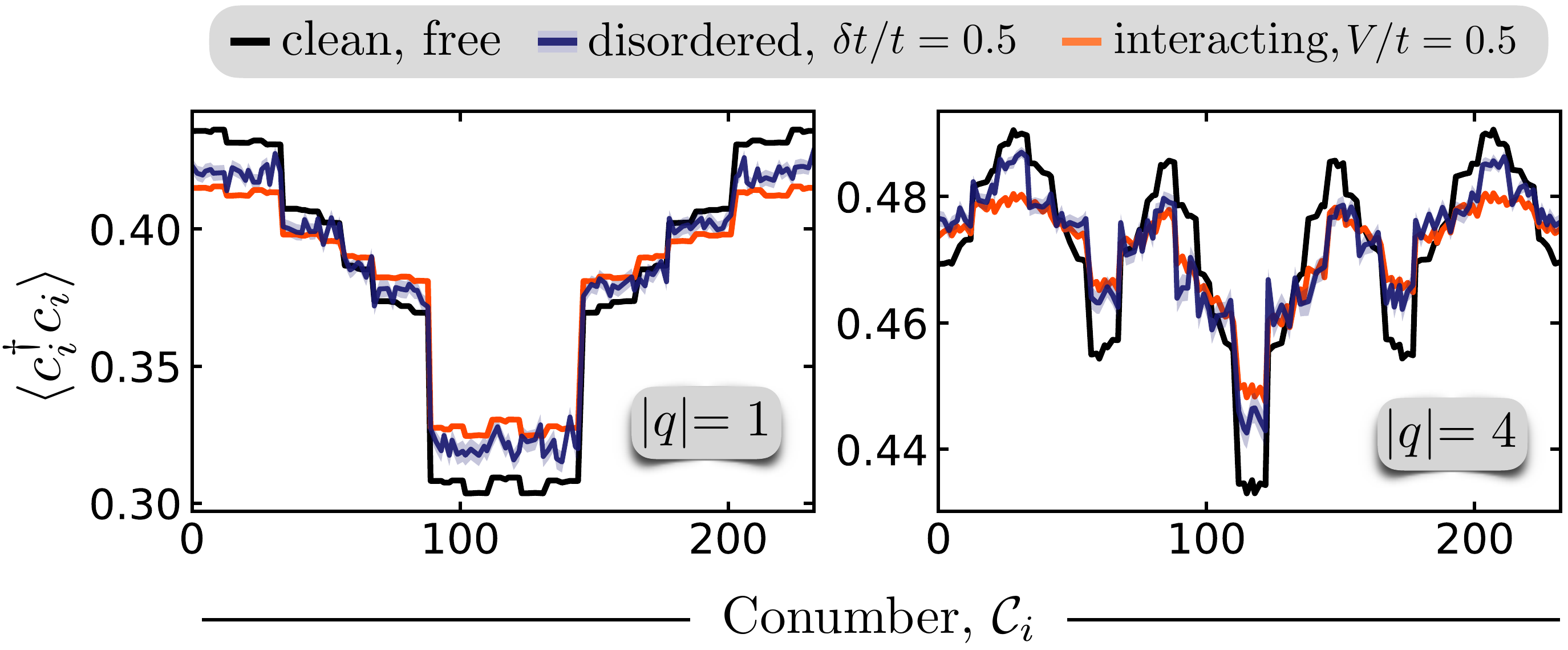} 
    \caption{The perpendicular space charge density for the clean, free (black), disordered (blue) and interacting (orange) Fibonacci chain. For the disordered case, $\delta t/t =0.5$, and the disorder average is taken over 1500 realizations. The light blue background indicates the statistical error to the mean.}
    \label{fig:Interacting+temperature}
\end{figure}

\section{Robustness}
We find that the observed oscillations are robust to moderate amounts of disorder, local interactions, as well as edge effects. To study the effect of disorder, we add off-diagonal noise to the Hamiltonian: $t_i \rightarrow t_i + \delta t_i$ with $\{\delta t_i\}$ a set of uniformly distributed random variables on the interval $[-\delta t, \delta t]$. The oscillations in a gap are expected to survive as long as the disorder doesn't close the gap \cite{jagannathanNonmonotonicCrossoverScaling2019}. The threshold disorder strength at which the oscillation breaks down therefore depends on the size of the gap, $\Delta_q$. An in-depth study of the relationship between the threshold disorder strength and the gap size deserves its own study. In this report, we comment on a few interesting observations. Fig.~\ref{fig:Interacting+temperature} depicts the disorder averaged result over 1500 realizations with disorder strength $\delta t/t = 0.5$ and fixed particle number for the cases $q=-1,4$. We find that the charge oscillations prevail even in regimes where the disorder strength exceeds the gap size. For instance, the oscillations in the $q=-1,4$ gaps are clearly visible even for $\delta t/t = 0.5$, while $\Delta_{-1}\sim 0.3$ and $\Delta_4 \sim 0.1$.
In Appendix~\ref{sec:Rob_app}, we show a similar figure for smaller gaps ($|q|=3,5$), where the loss of the $q$-oscillations can be seen with increasing disorder strength (see Fig.~\ref{fig:disorder_35}). We further study the effect of nearest neighbor repulsion of the form $V c^\dagger_ic^\dagger_{i+1} c_{i+1} c_i $ using the density-matrix renormalization group (DMRG)~\cite{SchollwoeckDMRG, tenpy}. As shown in Fig.~\ref{fig:Interacting+temperature}, the oscillation persists with a renormalized amplitude similar to the disordered case. Open boundary conditions introduce deviations to the charge density for a small number of sites close to the edges, but since sites close to the edge in real space are uniformly distributed in perpendicular space, these sites can be safely ignored without changing the Fourier amplitudes in a significant way. 


\section{Entanglement entropy oscillations in perpendicular space} 
Similar oscillations are also present in the entanglement entropy.
%
The von Neumann entanglement entropy (EE) of subsystem $A_{\ell} = [1,...,\ell]$ with $B_{\ell}=[\ell+1, ..., N]$ is given by
\begin{equation}
     S_{\ell} = S(\rho_{A_{\ell}}) =  -\text{Tr} \left[ \text{ln}(\rho_{A_{\ell}}) \rho_{A_{\ell}} \right].
\end{equation}
In quantum critical gapless systems, the von Neumann entanglement entropy follows the Cardy-Calabrese~\cite{Calabrese_2004} formula. Corrections to this scaling behavior for the Fibonacci chain have been calculated \cite{igloiEntanglementEntropyAperiodic2007} with the Fibonacci modulation treated as an irrelevant perturbation. The situation is a little bit different if the perturbation is relevant. In gapped phases, the von Neumann entropy spectrum saturates in the bulk. If the gaps are topologically non-trivial, strong parity effects appear in the entanglement spectrum that persist even deep in the bulk. This alternating bulk entanglement entropy has been shown to classify the topological phase of one-dimensional quantum $t_1$-$t_2$ (SSH model) and $t_1$-$t_2$-$t_3$ chains~\cite{Chunyu}. 
\begin{figure}
    \centering
    \includegraphics[width=\columnwidth]{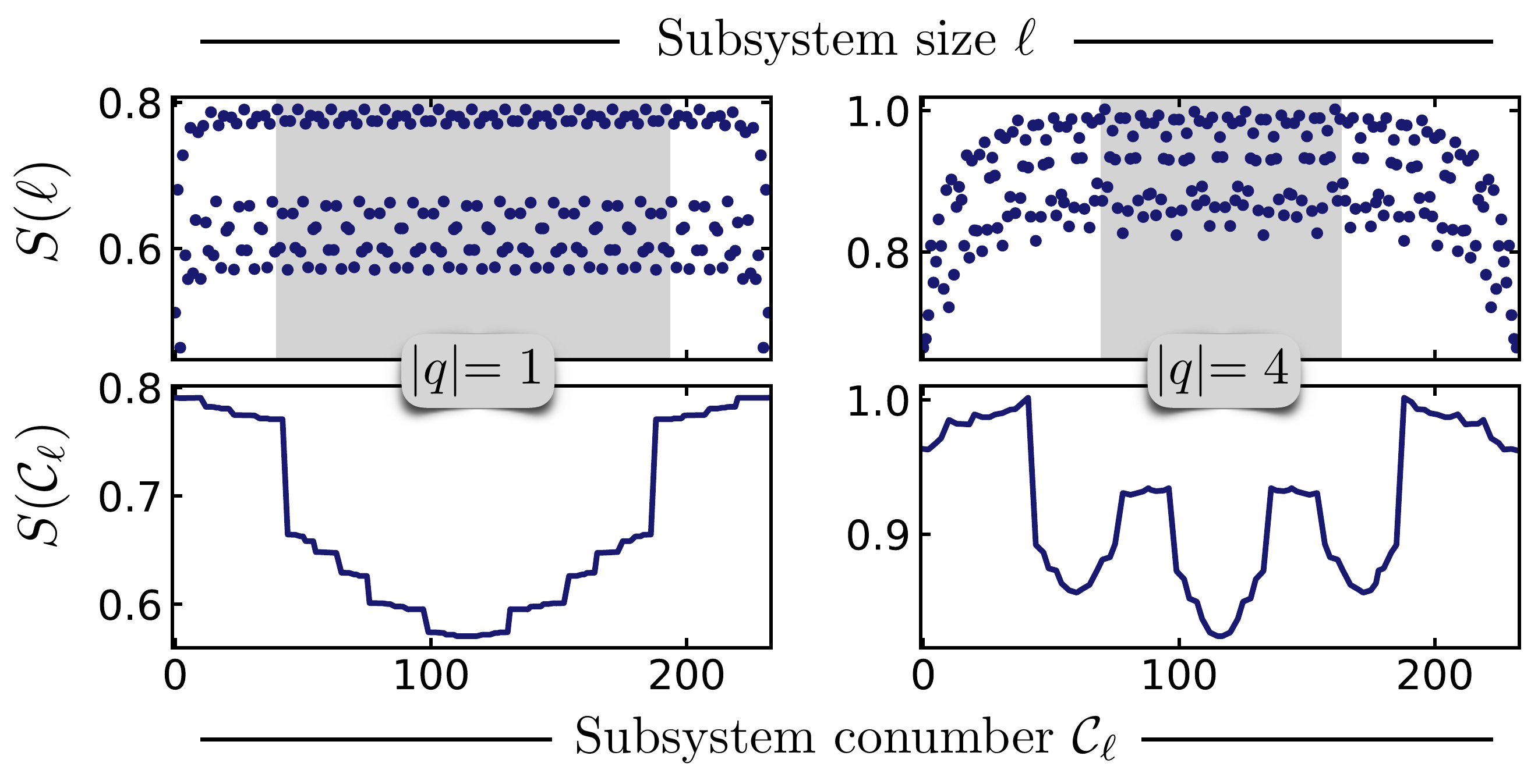}
    \caption{The entanglement entropy, $S_{\ell}$, in real (upper row) and co-number (lower row) space in gaps with labels $q=-1$ and $q=4$. We only map data points from the bulk (marked by a shaded background) to perpendicular space.}
    \label{fig:ent_oscillations}
\end{figure}
A generalization of this behavior can be seen in the entanglement entropy (EE) of a Fibonacci approximant by using the conumber transform. The subsystem $A_\ell (B_\ell)$ is defined as all sites to the left (right) of the bond between the $\ell$th and $(\ell+1)$th site. The EE of a Fibonacci approximant as a function of the cut location $\ell$ is shown in the upper row of Fig.~\ref{fig:ent_oscillations} for two different fillings. Here, we have applied open boundary conditions. The bottom row shows the same data for the EE, however plotted versus the co-number of the $\ell$th site---the site immediately to the left of the cut. In the lower plots, only the data for sites in the interior of the chain (the grey zone of the upper plot) are shown, since open boundaries introduce spurious edge effects. Just like the fluctuations presented above for the charge density, the bulk entanglement spectrum oscillates between particular values when the chemical potential is tuned inside a gap. These fluctuations map to full-period oscillations in perpendicular space, where the number of oscillations predicts precisely the topological label of the corresponding gap. The alternating EE reported in~\cite{Chunyu} for the $t_1$-$t_2$ model, which is the lowest order approximant of the Fibonacci chain, can be interpreted as corresponding to $|q|=1$ gap. Here, \emph{perpendicular space} only contains two in-equivalent points (corresponding to A and B sites), and the entanglement alternates between two values in the bulk. This can be interpreted by the formation of a valence bond at dimers in the chain in the large modulation limit. When analyzing larger approximants, new higher labelled gaps open in the single particle spectrum, leading to more complex entanglement structures which ultimately lead to the appearance of higher order oscillations---similar to the results and RG analysis presented previously for the charge density. A detailed analysis of the ground state entanglement is subject of an upcoming work.

\section{Conclusions}
We have shown that studying observables such as the charge density or the entanglement entropy in perpendicular space reveals the topological information stored in the many-body wavefunction of the Fibonacci chain. In particular, the gap labels of the Fibonacci chain can be measured by observing the oscillations of the charge density in perpendicular space when the filling is tuned to its topological gaps. An interesting direction for future work, finally, concerns the generalization of our results to quasicrystals in higher dimensions, which are characterized by several topological indices.





\begin{thebibliography}{30}%
\makeatletter
\providecommand \@ifxundefined [1]{%
 \@ifx{#1\undefined}
}%
\providecommand \@ifnum [1]{%
 \ifnum #1\expandafter \@firstoftwo
 \else \expandafter \@secondoftwo
 \fi
}%
\providecommand \@ifx [1]{%
 \ifx #1\expandafter \@firstoftwo
 \else \expandafter \@secondoftwo
 \fi
}%
\providecommand \natexlab [1]{#1}%
\providecommand \enquote  [1]{``#1''}%
\providecommand \bibnamefont  [1]{#1}%
\providecommand \bibfnamefont [1]{#1}%
\providecommand \citenamefont [1]{#1}%
\providecommand \href@noop [0]{\@secondoftwo}%
\providecommand \href [0]{\begingroup \@sanitize@url \@href}%
\providecommand \@href[1]{\@@startlink{#1}\@@href}%
\providecommand \@@href[1]{\endgroup#1\@@endlink}%
\providecommand \@sanitize@url [0]{\catcode `\\12\catcode `\$12\catcode
  `\&12\catcode `\#12\catcode `\^12\catcode `\_12\catcode `\%12\relax}%
\providecommand \@@startlink[1]{}%
\providecommand \@@endlink[0]{}%
\providecommand \url  [0]{\begingroup\@sanitize@url \@url }%
\providecommand \@url [1]{\endgroup\@href {#1}{\urlprefix }}%
\providecommand \urlprefix  [0]{URL }%
\providecommand \Eprint [0]{\href }%
\providecommand \doibase [0]{http://dx.doi.org/}%
\providecommand \selectlanguage [0]{\@gobble}%
\providecommand \bibinfo  [0]{\@secondoftwo}%
\providecommand \bibfield  [0]{\@secondoftwo}%
\providecommand \translation [1]{[#1]}%
\providecommand \BibitemOpen [0]{}%
\providecommand \bibitemStop [0]{}%
\providecommand \bibitemNoStop [0]{.\EOS\space}%
\providecommand \EOS [0]{\spacefactor3000\relax}%
\providecommand \BibitemShut  [1]{\csname bibitem#1\endcsname}%
\let\auto@bib@innerbib\@empty
\bibitem [{\citenamefont {Kraus}\ and\ \citenamefont
  {Zilberberg}(2016)}]{krausQuasiperiodicityTopologyTranscend2016}%
  \BibitemOpen
  \bibfield  {author} {\bibinfo {author} {\bibfnamefont {Y.~E.}\ \bibnamefont
  {Kraus}}\ and\ \bibinfo {author} {\bibfnamefont {O.}~\bibnamefont
  {Zilberberg}},\ }\href {\doibase 10.1038/nphys3784} {\bibfield  {journal}
  {\bibinfo  {journal} {Nature Physics}\ }\textbf {\bibinfo {volume} {12}},\
  \bibinfo {pages} {624} (\bibinfo {year} {2016})}\BibitemShut {NoStop}%
\bibitem [{\citenamefont {Flicker}\ and\ \citenamefont {van
  Wezel}(2015)}]{flickerQuasiperiodicity2DTopology2015}%
  \BibitemOpen
  \bibfield  {author} {\bibinfo {author} {\bibfnamefont {F.}~\bibnamefont
  {Flicker}}\ and\ \bibinfo {author} {\bibfnamefont {J.}~\bibnamefont {van
  Wezel}},\ }\href {\doibase 10.1209/0295-5075/111/37008} {\bibfield  {journal}
  {\bibinfo  {journal} {EPL (Europhysics Letters)}\ }\textbf {\bibinfo {volume}
  {111}},\ \bibinfo {pages} {37008} (\bibinfo {year} {2015})}\BibitemShut
  {NoStop}%
\bibitem [{\citenamefont {Bellissard}\ \emph {et~al.}(1992)\citenamefont
  {Bellissard}, \citenamefont {Bovier},\ and\ \citenamefont
  {Ghez}}]{bellissardGapLabellingTheorems1992}%
  \BibitemOpen
  \bibfield  {author} {\bibinfo {author} {\bibfnamefont {J.}~\bibnamefont
  {Bellissard}}, \bibinfo {author} {\bibfnamefont {A.}~\bibnamefont {Bovier}},
  \ and\ \bibinfo {author} {\bibfnamefont {A.~M.}\ \bibnamefont {Ghez}},\
  }\href@noop {} {\bibfield  {journal} {\bibinfo  {journal} {Reviews in
  Mathematical Physics}\ } (\bibinfo {year} {1992})}\BibitemShut {NoStop}%
\bibitem [{\citenamefont
  {Jagannathan}(2020)}]{jagannathanFibonacciQuasicrystalCase2020}%
  \BibitemOpen
  \bibfield  {author} {\bibinfo {author} {\bibfnamefont {A.}~\bibnamefont
  {Jagannathan}},\ }\href {http://arxiv.org/abs/2012.14744} {\bibfield
  {journal} {\bibinfo  {journal} {arXiv:2012.14744 [cond-mat]}\ } (\bibinfo
  {year} {2020})},\ \bibinfo {note} {arXiv: 2012.14744}\BibitemShut {NoStop}%
\bibitem [{\citenamefont {Kraus}\ and\ \citenamefont
  {Zilberberg}(2012)}]{krausTopologicalEquivalenceFibonacci2012}%
  \BibitemOpen
  \bibfield  {author} {\bibinfo {author} {\bibfnamefont {Y.~E.}\ \bibnamefont
  {Kraus}}\ and\ \bibinfo {author} {\bibfnamefont {O.}~\bibnamefont
  {Zilberberg}},\ }\href {\doibase 10.1103/PhysRevLett.109.116404} {\bibfield
  {journal} {\bibinfo  {journal} {Physical Review Letters}\ }\textbf {\bibinfo
  {volume} {109}},\ \bibinfo {pages} {116404} (\bibinfo {year}
  {2012})}\BibitemShut {NoStop}%
\bibitem [{\citenamefont {Baboux}\ \emph {et~al.}(2017)\citenamefont {Baboux},
  \citenamefont {Levy}, \citenamefont {Lema{\^i}tre}, \citenamefont
  {G{\'o}mez}, \citenamefont {Galopin}, \citenamefont {Le~Gratiet},
  \citenamefont {Sagnes}, \citenamefont {Amo}, \citenamefont {Bloch},\ and\
  \citenamefont {Akkermans}}]{babouxMeasuringTopologicalInvariants2017}%
  \BibitemOpen
  \bibfield  {author} {\bibinfo {author} {\bibfnamefont {F.}~\bibnamefont
  {Baboux}}, \bibinfo {author} {\bibfnamefont {E.}~\bibnamefont {Levy}},
  \bibinfo {author} {\bibfnamefont {A.}~\bibnamefont {Lema{\^i}tre}}, \bibinfo
  {author} {\bibfnamefont {C.}~\bibnamefont {G{\'o}mez}}, \bibinfo {author}
  {\bibfnamefont {E.}~\bibnamefont {Galopin}}, \bibinfo {author} {\bibfnamefont
  {L.}~\bibnamefont {Le~Gratiet}}, \bibinfo {author} {\bibfnamefont
  {I.}~\bibnamefont {Sagnes}}, \bibinfo {author} {\bibfnamefont
  {A.}~\bibnamefont {Amo}}, \bibinfo {author} {\bibfnamefont {J.}~\bibnamefont
  {Bloch}}, \ and\ \bibinfo {author} {\bibfnamefont {E.}~\bibnamefont
  {Akkermans}},\ }\href {\doibase 10.1103/PhysRevB.95.161114} {\bibfield
  {journal} {\bibinfo  {journal} {Physical Review B}\ }\textbf {\bibinfo
  {volume} {95}},\ \bibinfo {pages} {161114(R)} (\bibinfo {year}
  {2017})}\BibitemShut {NoStop}%
\bibitem [{\citenamefont {Verbin}\ \emph {et~al.}(2015)\citenamefont {Verbin},
  \citenamefont {Zilberberg}, \citenamefont {Lahini}, \citenamefont {Kraus},\
  and\ \citenamefont {Silberberg}}]{verbinTopologicalPumpingPhotonic2015}%
  \BibitemOpen
  \bibfield  {author} {\bibinfo {author} {\bibfnamefont {M.}~\bibnamefont
  {Verbin}}, \bibinfo {author} {\bibfnamefont {O.}~\bibnamefont {Zilberberg}},
  \bibinfo {author} {\bibfnamefont {Y.}~\bibnamefont {Lahini}}, \bibinfo
  {author} {\bibfnamefont {Y.~E.}\ \bibnamefont {Kraus}}, \ and\ \bibinfo
  {author} {\bibfnamefont {Y.}~\bibnamefont {Silberberg}},\ }\href {\doibase
  10.1103/PhysRevB.91.064201} {\bibfield  {journal} {\bibinfo  {journal}
  {Physical Review B}\ }\textbf {\bibinfo {volume} {91}},\ \bibinfo {pages}
  {064201} (\bibinfo {year} {2015})}\BibitemShut {NoStop}%
\bibitem [{\citenamefont {Rai}\ \emph {et~al.}(2019)\citenamefont {Rai},
  \citenamefont {Haas},\ and\ \citenamefont
  {Jagannathan}}]{raiProximityEffectSuperconductorquasicrystal2019}%
  \BibitemOpen
  \bibfield  {author} {\bibinfo {author} {\bibfnamefont {G.}~\bibnamefont
  {Rai}}, \bibinfo {author} {\bibfnamefont {S.}~\bibnamefont {Haas}}, \ and\
  \bibinfo {author} {\bibfnamefont {A.}~\bibnamefont {Jagannathan}},\ }\href
  {\doibase 10.1103/PhysRevB.100.165121} {\bibfield  {journal} {\bibinfo
  {journal} {Physical Review B}\ }\textbf {\bibinfo {volume} {100}},\ \bibinfo
  {pages} {165121} (\bibinfo {year} {2019})}\BibitemShut {NoStop}%
\bibitem [{\citenamefont {Dareau}\ \emph {et~al.}(2017)\citenamefont {Dareau},
  \citenamefont {Levy}, \citenamefont {Aguilera}, \citenamefont {Bouganne},
  \citenamefont {Akkermans}, \citenamefont {Gerbier},\ and\ \citenamefont
  {Beugnon}}]{dareauRevealingTopologyQuasicrystals2017}%
  \BibitemOpen
  \bibfield  {author} {\bibinfo {author} {\bibfnamefont {A.}~\bibnamefont
  {Dareau}}, \bibinfo {author} {\bibfnamefont {E.}~\bibnamefont {Levy}},
  \bibinfo {author} {\bibfnamefont {M.~B.}\ \bibnamefont {Aguilera}}, \bibinfo
  {author} {\bibfnamefont {R.}~\bibnamefont {Bouganne}}, \bibinfo {author}
  {\bibfnamefont {E.}~\bibnamefont {Akkermans}}, \bibinfo {author}
  {\bibfnamefont {F.}~\bibnamefont {Gerbier}}, \ and\ \bibinfo {author}
  {\bibfnamefont {J.}~\bibnamefont {Beugnon}},\ }\href {\doibase
  10.1103/PhysRevLett.119.215304} {\bibfield  {journal} {\bibinfo  {journal}
  {Physical Review Letters}\ }\textbf {\bibinfo {volume} {119}},\ \bibinfo
  {pages} {215304} (\bibinfo {year} {2017})},\ \bibinfo {note} {arXiv:
  1607.00901}\BibitemShut {NoStop}%
\bibitem [{\citenamefont {Sire}\ and\ \citenamefont
  {Mosseri}(1990)}]{sireExcitationSpectrumExtended1990}%
  \BibitemOpen
  \bibfield  {author} {\bibinfo {author} {\bibfnamefont {C.}~\bibnamefont
  {Sire}}\ and\ \bibinfo {author} {\bibfnamefont {R.}~\bibnamefont {Mosseri}},\
  }\href {\doibase 10.1051/jphys:0199000510150156900} {\bibfield  {journal}
  {\bibinfo  {journal} {Journal de Physique}\ }\textbf {\bibinfo {volume}
  {51}},\ \bibinfo {pages} {1569} (\bibinfo {year} {1990})}\BibitemShut
  {NoStop}%
\bibitem [{\citenamefont {Kempkes}\ \emph {et~al.}(2019)\citenamefont
  {Kempkes}, \citenamefont {Slot}, \citenamefont {Freeney}, \citenamefont
  {Zevenhuizen}, \citenamefont {Vanmaekelbergh}, \citenamefont {Swart},\ and\
  \citenamefont {Smith}}]{kempkesDesignCharacterizationElectrons2019}%
  \BibitemOpen
  \bibfield  {author} {\bibinfo {author} {\bibfnamefont {S.~N.}\ \bibnamefont
  {Kempkes}}, \bibinfo {author} {\bibfnamefont {M.}~\bibnamefont {Slot}},
  \bibinfo {author} {\bibfnamefont {S.~E.}\ \bibnamefont {Freeney}}, \bibinfo
  {author} {\bibfnamefont {S.~J.~M.}\ \bibnamefont {Zevenhuizen}}, \bibinfo
  {author} {\bibfnamefont {D.}~\bibnamefont {Vanmaekelbergh}}, \bibinfo
  {author} {\bibfnamefont {I.}~\bibnamefont {Swart}}, \ and\ \bibinfo {author}
  {\bibfnamefont {C.~M.}\ \bibnamefont {Smith}},\ }\href {\doibase
  10.1038/s41567-018-0328-0} {\bibfield  {journal} {\bibinfo  {journal} {Nature
  physics}\ }\textbf {\bibinfo {volume} {15}},\ \bibinfo {pages} {127}
  (\bibinfo {year} {2019})}\BibitemShut {NoStop}%
\bibitem [{\citenamefont {Slot}\ \emph {et~al.}(2017)\citenamefont {Slot},
  \citenamefont {Gardenier}, \citenamefont {Jacobse}, \citenamefont {van
  Miert}, \citenamefont {Kempkes}, \citenamefont {Zevenhuizen}, \citenamefont
  {Smith}, \citenamefont {Vanmaekelbergh},\ and\ \citenamefont
  {Swart}}]{slotExperimentalRealizationCharacterization2017}%
  \BibitemOpen
  \bibfield  {author} {\bibinfo {author} {\bibfnamefont {M.~R.}\ \bibnamefont
  {Slot}}, \bibinfo {author} {\bibfnamefont {T.~S.}\ \bibnamefont {Gardenier}},
  \bibinfo {author} {\bibfnamefont {P.~H.}\ \bibnamefont {Jacobse}}, \bibinfo
  {author} {\bibfnamefont {G.~C.~P.}\ \bibnamefont {van Miert}}, \bibinfo
  {author} {\bibfnamefont {S.~N.}\ \bibnamefont {Kempkes}}, \bibinfo {author}
  {\bibfnamefont {S.~J.~M.}\ \bibnamefont {Zevenhuizen}}, \bibinfo {author}
  {\bibfnamefont {C.~M.}\ \bibnamefont {Smith}}, \bibinfo {author}
  {\bibfnamefont {D.}~\bibnamefont {Vanmaekelbergh}}, \ and\ \bibinfo {author}
  {\bibfnamefont {I.}~\bibnamefont {Swart}},\ }\href {\doibase
  10.1038/nphys4105} {\bibfield  {journal} {\bibinfo  {journal} {Nature
  Physics}\ }\textbf {\bibinfo {volume} {13}},\ \bibinfo {pages} {672}
  (\bibinfo {year} {2017})}\BibitemShut {NoStop}%
\bibitem [{\citenamefont {Tanese}\ \emph {et~al.}(2014)\citenamefont {Tanese},
  \citenamefont {Gurevich}, \citenamefont {Baboux}, \citenamefont {Jacqmin},
  \citenamefont {Lema{\^i}tre}, \citenamefont {Galopin}, \citenamefont
  {Sagnes}, \citenamefont {Amo}, \citenamefont {Bloch},\ and\ \citenamefont
  {Akkermans}}]{taneseFractalEnergySpectrum2014}%
  \BibitemOpen
  \bibfield  {author} {\bibinfo {author} {\bibfnamefont {D.}~\bibnamefont
  {Tanese}}, \bibinfo {author} {\bibfnamefont {E.}~\bibnamefont {Gurevich}},
  \bibinfo {author} {\bibfnamefont {F.}~\bibnamefont {Baboux}}, \bibinfo
  {author} {\bibfnamefont {T.}~\bibnamefont {Jacqmin}}, \bibinfo {author}
  {\bibfnamefont {A.}~\bibnamefont {Lema{\^i}tre}}, \bibinfo {author}
  {\bibfnamefont {E.}~\bibnamefont {Galopin}}, \bibinfo {author} {\bibfnamefont
  {I.}~\bibnamefont {Sagnes}}, \bibinfo {author} {\bibfnamefont
  {A.}~\bibnamefont {Amo}}, \bibinfo {author} {\bibfnamefont {J.}~\bibnamefont
  {Bloch}}, \ and\ \bibinfo {author} {\bibfnamefont {E.}~\bibnamefont
  {Akkermans}},\ }\href {\doibase 10.1103/PhysRevLett.112.146404} {\bibfield
  {journal} {\bibinfo  {journal} {Physical Review Letters}\ }\textbf {\bibinfo
  {volume} {112}},\ \bibinfo {pages} {146404} (\bibinfo {year}
  {2014})}\BibitemShut {NoStop}%
\bibitem [{Note1()}]{Note1}%
  \BibitemOpen
  \bibinfo {note} {We pick $t_A$ and $t_B$ such that the average hopping $t =
  \protect \frac {F_{n-1}t_A + F_{n-2}t_B}{F_n} = 1$.}\BibitemShut {Stop}%
\bibitem [{Note2()}]{Note2}%
  \BibitemOpen
  \bibinfo {note} {We used the Python package Kwant \cite
  {grothKwantSoftwarePackage2014} to set up the Hamiltonian matrix}\BibitemShut
  {NoStop}%
\bibitem [{\citenamefont {Sire}\ and\ \citenamefont
  {Mosseri}(1989)}]{sireSpectrum1DQuasicrystals1989}%
  \BibitemOpen
  \bibfield  {author} {\bibinfo {author} {\bibfnamefont {C.}~\bibnamefont
  {Sire}}\ and\ \bibinfo {author} {\bibfnamefont {R.}~\bibnamefont {Mosseri}},\
  }\href {\doibase 10.1051/jphys:0198900500240344700} {\bibfield  {journal}
  {\bibinfo  {journal} {Journal de Physique}\ }\textbf {\bibinfo {volume}
  {50}},\ \bibinfo {pages} {3447} (\bibinfo {year} {1989})}\BibitemShut
  {NoStop}%
\bibitem [{Note3()}]{Note3}%
  \BibitemOpen
  \bibinfo {note} {$\protect \hat {\protect \mathcal {H}} = \protect \hat
  {\protect \mathcal {H}}_0 + \protect \hat {\protect \mathcal {H}}_w$ itself
  has a simpler form in perpendicular space. It is a symmetric Toeplitz matrix
  with with 4 filled diagonals (other than the chemical in the main
  diagonal)}\BibitemShut {NoStop}%
\bibitem [{\citenamefont {Kalugin}\ \emph {et~al.}(1986)\citenamefont
  {Kalugin}, \citenamefont {Kitaev},\ and\ \citenamefont
  {Levitov}}]{kaluginElectronSpectrumOnedimensional1986}%
  \BibitemOpen
  \bibfield  {author} {\bibinfo {author} {\bibfnamefont {P.~A.}\ \bibnamefont
  {Kalugin}}, \bibinfo {author} {\bibfnamefont {A.~Y.}\ \bibnamefont {Kitaev}},
  \ and\ \bibinfo {author} {\bibfnamefont {L.~S.}\ \bibnamefont {Levitov}},\
  }\href@noop {} {\bibfield  {journal} {\bibinfo  {journal} {Journal of
  Experimental and Theoretical Physics}\ }\textbf {\bibinfo {volume} {64}},\
  \bibinfo {pages} {6} (\bibinfo {year} {1986})}\BibitemShut {NoStop}%
\bibitem [{\citenamefont {Niu}\ and\ \citenamefont
  {Nori}(1986)}]{niuRenormalizationGroupStudyOneDimensional1986}%
  \BibitemOpen
  \bibfield  {author} {\bibinfo {author} {\bibfnamefont {Q.}~\bibnamefont
  {Niu}}\ and\ \bibinfo {author} {\bibfnamefont {F.}~\bibnamefont {Nori}},\
  }\href {\doibase 10.1103/PhysRevLett.57.2057} {\bibfield  {journal} {\bibinfo
   {journal} {Physical Review Letters}\ }\textbf {\bibinfo {volume} {57}},\
  \bibinfo {pages} {2057} (\bibinfo {year} {1986})}\BibitemShut {NoStop}%
\bibitem [{Note4()}]{Note4}%
  \BibitemOpen
  \bibinfo {note} {We use the convention $F_1 = 1$ and $F_2 = 2$}\BibitemShut
  {NoStop}%
\bibitem [{\citenamefont
  {Brown}(1964)}]{brownZeckendorfTheoremAppucations1964}%
  \BibitemOpen
  \bibfield  {author} {\bibinfo {author} {\bibfnamefont {J.~L.}\ \bibnamefont
  {Brown}},\ }\href@noop {} {\bibfield  {journal} {\bibinfo  {journal} {The
  Fibonacci Quarterly}\ }\textbf {\bibinfo {volume} {2}} (\bibinfo {year}
  {1964})}\BibitemShut {NoStop}%
\bibitem [{\citenamefont {Jagannathan}\ \emph {et~al.}(2019)\citenamefont
  {Jagannathan}, \citenamefont {Jeena},\ and\ \citenamefont
  {Tarzia}}]{jagannathanNonmonotonicCrossoverScaling2019}%
  \BibitemOpen
  \bibfield  {author} {\bibinfo {author} {\bibfnamefont {A.}~\bibnamefont
  {Jagannathan}}, \bibinfo {author} {\bibfnamefont {P.}~\bibnamefont {Jeena}},
  \ and\ \bibinfo {author} {\bibfnamefont {M.}~\bibnamefont {Tarzia}},\ }\href
  {\doibase 10.1103/PhysRevB.99.054203} {\bibfield  {journal} {\bibinfo
  {journal} {Physical Review B}\ }\textbf {\bibinfo {volume} {99}},\ \bibinfo
  {pages} {054203} (\bibinfo {year} {2019})}\BibitemShut {NoStop}%
\bibitem [{\citenamefont {Schollw\"{o}ck}(2011)}]{SchollwoeckDMRG}%
  \BibitemOpen
  \bibfield  {author} {\bibinfo {author} {\bibfnamefont {U.}~\bibnamefont
  {Schollw\"{o}ck}},\ }\href {\doibase
  https://doi.org/10.1016/j.aop.2010.09.012} {\bibfield  {journal} {\bibinfo
  {journal} {Annals of Physics}\ }\textbf {\bibinfo {volume} {326}},\ \bibinfo
  {pages} {96} (\bibinfo {year} {2011})},\ \bibinfo {note} {january 2011
  Special Issue}\BibitemShut {NoStop}%
\bibitem [{\citenamefont {Hauschild}\ and\ \citenamefont
  {Pollmann}(2018)}]{tenpy}%
  \BibitemOpen
  \bibfield  {author} {\bibinfo {author} {\bibfnamefont {J.}~\bibnamefont
  {Hauschild}}\ and\ \bibinfo {author} {\bibfnamefont {F.}~\bibnamefont
  {Pollmann}},\ }\href {\doibase 10.21468/SciPostPhysLectNotes.5} {\bibfield
  {journal} {\bibinfo  {journal} {SciPost Phys. Lect. Notes}\ } (\bibinfo
  {year} {2018}),\ 10.21468/SciPostPhysLectNotes.5},\ \bibinfo {note} {code
  available from \url{https://github.com/tenpy/tenpy}},\ \Eprint
  {http://arxiv.org/abs/1805.00055} {arXiv:1805.00055} \BibitemShut {NoStop}%
\bibitem [{\citenamefont {Calabrese}\ and\ \citenamefont
  {Cardy}(2004)}]{Calabrese_2004}%
  \BibitemOpen
  \bibfield  {author} {\bibinfo {author} {\bibfnamefont {P.}~\bibnamefont
  {Calabrese}}\ and\ \bibinfo {author} {\bibfnamefont {J.}~\bibnamefont
  {Cardy}},\ }\href {\doibase 10.1088/1742-5468/2004/06/p06002} {\bibfield
  {journal} {\bibinfo  {journal} {Journal of Statistical Mechanics: Theory and
  Experiment}\ }\textbf {\bibinfo {volume} {2004}},\ \bibinfo {pages} {P06002}
  (\bibinfo {year} {2004})}\BibitemShut {NoStop}%
\bibitem [{\citenamefont {Igl{\'o}i}\ \emph {et~al.}(2007)\citenamefont
  {Igl{\'o}i}, \citenamefont {Juh{\'a}sz},\ and\ \citenamefont
  {Zimbor{\'a}s}}]{igloiEntanglementEntropyAperiodic2007}%
  \BibitemOpen
  \bibfield  {author} {\bibinfo {author} {\bibfnamefont {F.}~\bibnamefont
  {Igl{\'o}i}}, \bibinfo {author} {\bibfnamefont {R.}~\bibnamefont
  {Juh{\'a}sz}}, \ and\ \bibinfo {author} {\bibfnamefont {Z.}~\bibnamefont
  {Zimbor{\'a}s}},\ }\href {\doibase 10.1209/0295-5075/79/37001} {\bibfield
  {journal} {\bibinfo  {journal} {Europhysics Letters (EPL)}\ }\textbf
  {\bibinfo {volume} {79}},\ \bibinfo {pages} {37001} (\bibinfo {year}
  {2007})}\BibitemShut {NoStop}%
\bibitem [{\citenamefont {Tan}\ \emph {et~al.}(2020)\citenamefont {Tan},
  \citenamefont {Saleur},\ and\ \citenamefont {Haas}}]{Chunyu}%
  \BibitemOpen
  \bibfield  {author} {\bibinfo {author} {\bibfnamefont {C.}~\bibnamefont
  {Tan}}, \bibinfo {author} {\bibfnamefont {H.}~\bibnamefont {Saleur}}, \ and\
  \bibinfo {author} {\bibfnamefont {S.}~\bibnamefont {Haas}},\ }\href {\doibase
  10.1103/PhysRevB.101.235155} {\bibfield  {journal} {\bibinfo  {journal}
  {Phys. Rev. B}\ }\textbf {\bibinfo {volume} {101}},\ \bibinfo {pages}
  {235155} (\bibinfo {year} {2020})}\BibitemShut {NoStop}%
\bibitem [{\citenamefont {Groth}\ \emph {et~al.}(2014)\citenamefont {Groth},
  \citenamefont {Wimmer}, \citenamefont {Akhmerov},\ and\ \citenamefont
  {Waintal}}]{grothKwantSoftwarePackage2014}%
  \BibitemOpen
  \bibfield  {author} {\bibinfo {author} {\bibfnamefont {C.~W.}\ \bibnamefont
  {Groth}}, \bibinfo {author} {\bibfnamefont {M.}~\bibnamefont {Wimmer}},
  \bibinfo {author} {\bibfnamefont {A.~R.}\ \bibnamefont {Akhmerov}}, \ and\
  \bibinfo {author} {\bibfnamefont {X.}~\bibnamefont {Waintal}},\ }\href
  {\doibase 10.1088/1367-2630/16/6/063065} {\bibfield  {journal} {\bibinfo
  {journal} {New Journal of Physics}\ }\textbf {\bibinfo {volume} {16}},\
  \bibinfo {pages} {063065} (\bibinfo {year} {2014})}\BibitemShut {NoStop}%
\bibitem [{Note5()}]{Note5}%
  \BibitemOpen
  \bibinfo {note} {This symmetry is broken when PBC are applied in odd-length
  chains but is restored in the limit of infinite system size}\BibitemShut
  {NoStop}%
\bibitem [{Note6()}]{Note6}%
  \BibitemOpen
  \bibinfo {note} {Using the convention that the eigenstate at the edge of the
  Brillouin zone with label $\mu $ has the wavenumber $\protect \frac {2\pi
  (\mu + 1/2)}{F_n}$}\BibitemShut {NoStop}%
\end{thebibliography}
%

\begin{center}
    \textbf{Appendix}
\end{center}
\appendix
\section{Spectral analysis using RG in the strong modulation limit}

In the main text, we presented a procedure that recovers the charge density profile in the strong modulation limit, using an RG scheme in real space. In this limit, the band structure and the wave functions are well described in terms of a perturbative RG scheme where the small parameter is the coupling ratio, $\rho=t_A/t_B \ll 1$. Here, we present an alternate equivalent formulation focusing on the splitting of the spectrum.
 
 For $\rho=0$ the chain breaks up into molecules ($m$) and atoms ($a$), and the spectrum consists of three levels that broaden into three bands as $\rho$ is increased. The RG theory gives the relationship between spectra of a long chain and those of shorter chains. Thus one can show that the energy spectrum of the $n$th approximant has a symmetric structure due to the particle-hole symmetry. \footnote{This symmetry is broken when PBC are applied in odd-length chains but is restored in the limit of infinite system size}. One gets three main groups of levels: the lowest $F_{n-2}$ $m$ levels being the so-called molecular bonding ($m_-$), the highest $F_{n-2}$ being antibonding $m_+$) levels. The central group of $F_{n-3}$ levels are so-called atomic ($a$) levels. Each of these can be subdivided into three groups and so on. 

For sites in the conumbering system, the same demarcation occurs within the interval $[1,F_n]$. The first set of $F_{n-2}$ sites (the left subinterval) are $m$ sites and their partners lie in the right subinterval. The $a$ sites lie in the middle subinterval. Filling up the lowest levels results in a charge of 0.5 per $m$ site, filling up all the $a$ levels results in a charge of 1 per $a$ site. 
The p-h symmetry ensures that, for half-filling, all sites are filled equally, with $n_i=0.5$.  

With this notation in place we can now discuss the way total charge is built up on each site for different fillings.
$N_e$ in the following denotes the number of (spinless electrons). One can restrict to the lower half of the spectrum since the upper half can be deduced by p-h symmetry.
Let us consider the cases where $N_e$ is a Fibonacci number, $N_e=F_{n-j}$, where $j=2,3,...$. For any $j$, it is easy to check that $F_{n-j-1}=\mathrm{Mod}[q F_{n-1},F_n]$ with $q=(-1)^j F_{j}$. In other words, all the levels of the lowest band are completely filled upto the gap of label $q$.
\begin{enumerate}
\item For $N_e=F_{n-2}$ ($q=-1$), the lowest band ($m_-$) is completely filled. To leading order, the total charge is $0.5$ on $m$ sites, and 0 on $a$ sites. Plotted as a function of conumber one has a plateau where $n_i = 0.5$ on the left and on the right, and a central region $n_i = 0$. This zero order picture is modified by higher order corrections in $\rho$, that lead to finer structures within each region.  
\item For $N_e=F_{n-4}$ ($q=-3$), the lowest sub-band of the lowest band is completely filled.  This implies that each of the plateaus again splits into two plateaus. The number of maxima (Nb.half-maxima at each edge) is $2^2-1=3$ as expected from the $q$ value for this filling. 
\item For $N_e=F_{n-3}$ ($q=2$) the lowest band is partly filled, leading to different charges on the two types of $m$ sites. The total onsite occupancy is 0.25 for the $mm$ sites and 0.5 for the $am$ sites to leading order. These values are modified by higher order corrections when $\rho$ is sslowly turned on.
\end{enumerate}
When $N_e$ is not a Fibonacci number, we employ Zeckendorf's theorem in an analogous way to the previous section. Consider for example the case $N_e=F_{n-2}+F_{n-5}$, where one can check that $q=4$. Now the lowest band being completely filled, all of the $m$ sites have a charge of $0.5$. In addition, the occupancy $n_i$ equal to 0.5 at lowest order at the $am$ atom sites (atom sites which become $m$ after an RG step) and their partners to the right. For finite $\rho$, the resulting profile has 4 maxima.
 
The above special cases show how oscillations of charge can be explained based on the simplified picture in the strong modulation limit. 
\section{Perturbative analysis in the weak modulation limit}
When the modulation is weak, $t_A/t_B\sim 1$, we can treat the Fibonacci modulation as a perturbation to the homogeneous chain \cite{sireExcitationSpectrumExtended1990, sireSpectrum1DQuasicrystals1989}. In real space, the Fibonacci hopping Hamiltonian splits into two parts,
\begin{align}
    \hat{H} =& \underbrace{- \sum_{i=1}^{F_n} t_A c^\dagger_i c_{i+1} + h.c.}_{\hat{H}_0}
    \underbrace{- \sum_{i=1}^{F_n} (t_i - t_A) c^\dagger_i c_{i+1} + h.c.}_{\hat{H}_w}
\end{align}
$\hat{H}_0$ is the Hamiltonian of a periodic chain with nearest neighbor hopping $t_A$, and $\hat{H}_w$ introduces the Fibonacci modulation. We use periodic boundary conditions, so the index $i$ (and $j$ in the following) are understood to wrap around modulo $F_n$. The matrix representation of the Fibonacci hopping Hamiltonian in perpendicular space, which we will denote with $\hat{\mathcal{H}}$ has a particularly pleasing form. The only non-zero hopping terms lie on four diagonals. The $-t_A$s go in the $\pm F_{n-2}$th diagonal and the $-t_B$s go in the $\pm F_{n-1}$th diagonal.

The decomposition of $\hat{\mathcal{H}}$ as a sum of a term that represents the homogeneous chain $\hat{\mathcal{H}}_0$ and the Fibonacci perturbation $\hat{\mathcal{H}}_w$ takes the following form:
\begin{align}
    \hat{\mathcal{H}}_0 &= -\sum_{j=1}^{F_n} t_A c_j^\dagger c_{j+F_{n-2}} + t_A c_j^\dagger c_{j+ F_{n-1}} + h.c. \label{eq:H0_perp:S}\\
    \hat{\mathcal{H}}_w &= -\sum_{j=1}^{F_{n-2}} 2w c_j^\dagger c_{j+F_{n-1}} + h.c. \label{Hw_perp:S}
\end{align}
where we have used $2w = t_A-t_B$. $\hat{\mathcal{H}}_w$ connects each of the leftmost $F_{n-2}$ sites to the site $F_{n-1}$ places to its right. $\hat{\mathcal{H}}_0$ is a circulant matrix  and $\hat{\mathcal{H}}_w$ is a Toeplitz matrix with only the $\pm F_{n-1}$th diagonals filled.

Before applying the Fibonacci modulation, consider the eigenstates of the unperturbed Hamiltonian. Recall that the eigenvalues of $\hat{H}_0$ are given by $E_m = -2t_{A}\cos\left(\frac{2\pi m}{F_{n}}\right)$ with the integer $m$ constrained by $0\leq m \leq \lfloor F_n/2\rfloor$. Almost all of these are doubly degenerate. If $F_n$ is odd, the $m=0$ eigenvalue is non-degenerate. If $F_n$ is even, then addditionally, the $m = \lfloor F_n/2 \rfloor$ eigenvalue is also  non-degenerate. The corresponding eigenstates take the form of plane waves with wavenumber $\frac{2\pi}{F_n}m$. 

Remarkably, the unperturbed Hamiltonian represented in perpendicular space is a circulant matrix. This means that it is also diagonalized by a Fourier transformation and the eigenstates, which are plane waves in real space, are also plane waves in perpendicular space. But the wavenumbers are shuffled or the wavenumber that corresponds to a plane wave with a given energy is different. In perpendicular space, the eigenstates are plane waves labelled by the integer $\mu = 0,1,\ldots \lfloor F_n/2\rfloor$. As in real space, the eigenstates are almost all doubly degenerate, and the eigenstate with label $\mu$ has the wavenumber $\frac{2\pi}{F_n}\mu$. However, the energy of this plane wave is $\mathcal{E}_\mu = -2t_A \cos\left(\frac{2\pi \mu F_{n-1}}{F_n}\right)$. It is straightforward to check that the $\mathcal{E}_\mu$ labelled by $\mu$ and the $E_m$ labelled by $m$ represent the same set of energies. They are simply ordered differently. The perpendicular space and real space wavenumbers can be related to each other (modulo $F_n$) by equating $E_m$ and $\mathcal{E}_\mu$
\begin{align}\label{eq:mcongruentmuApp}
    m \equiv \mu F_{n-1} \pmod{F_n}
\end{align}
Bearing in mind that we will shortly introduce a perturbation with an $F_n$-site supercell, we describe the band structure of $\hat{H}_0$ as consisting of $F_n$ bands in a Brillouin zone that spans the interval $k\in\left[-\frac{\pi}{F_n},\frac{\pi}{F_n}\right[$. For the unperturbed problem, all the bands touch each other, reflecting the fact that we have done nothing other than downfold the usual $\cos\left(\frac{2 \pi k}{F_n}\right)$ dispersion onto a smaller Brillouin zone. Introducing the perturbation $\hat{\mathcal{H}}_w$ splits the doubly degenerate states of $\hat{\mathcal{H}}_0$, and opens $F_n - 1$ gaps. The gap opening can be seen using first-order perturbation theory. The edges of gaps with an even (odd) number of states below them are at $k = 0\left(\frac{\pi}{F_n}\right)$. We will show the calculation for the even case in detail. The odd case is entirely analogous.

The rank 2 degenerate subspace labelled by $\mu$ has two states $\ket{\mu, \pm}$ with energy $-2t_A\cos\left(\frac{2\pi F_{n-1}}{F_n}\mu\right)$. The first order
corrections are given by the eigensystem of a $2\times 2$ matrix $\hat{V}$ whose entries are $V_{\alpha\beta} = \bra{\mu,\alpha}\hat{\mathcal{H}}_w\ket{\mu,\beta}$ with $\alpha,\beta = \pm$.
\begin{align}
    V_{\alpha\beta} = \frac{-4w}{F_{n}}\begin{pmatrix}
F_{n-2}\cos\left( 2\pi\frac{F_{n-1} \mu}{F_{n}}\right) & \frac{\sin\left( 2\pi\frac{\mu}{F_{n}} F_{n-2}\right)}{\sin\left( 2\pi\frac{\mu}{F_{n}}\right)}\\
\frac{\sin\left( 2\pi\frac{\mu}{F_{n}} F_{n-2}\right)}{\sin\left( 2\pi\frac{\mu}{F_{n}}\right)} & F_{n-2}\cos\left( 2\pi\frac{F_{n-1} \mu}{F_{n}}\right)
\end{pmatrix}
\end{align}
The eigenvalues of $\displaystyle V_{\alpha\beta}$ are $\displaystyle \epsilon ^{\pm } =\frac{-4w}{F_{n}}\left[ F_{n-2}\cos\left( 2\pi\frac{F_{n-1} \mu}{F_{n}}\right) \pm \frac{\sin\left( 2\pi\frac{\mu}{F_{n}} F_{n-2}\right)}{\sin\left( 2\pi\frac{\mu}{F_{n}}\right)}\right]$, and the corresponding eigenvectors are $\displaystyle \left(\frac{1}{\sqrt{2}} ,\pm \ \frac{1}{\sqrt{2}}\right) \ $. So, we have that the plane waves with the wavenumber $\frac{2\pi \mu}{F_{n}}$ split into the states $\ket{\mu,\pm}^*$ with the energies $\mathcal{E}^{\pm}_\mu$
\begin{align}
    \mathcal{E}_\mu^{\pm} =& -\left( 2t_{A} + 4w\frac{F_{n-2}}{F_{n}}\right) \cos\left(\frac{2\pi F_{n-1}}{F_{n}}\mu\right) \nonumber\\
    &\mp \frac{4w}{F_{n}} \frac{\sin\left( 2\pi \frac{F_{n-2}}{F_{n}}\mu\right)}{\sin\left( \frac{2\pi }{F_{n}}\mu\right)}\label{eq:pert_eps:S}\\
    \braket{j|\mu,+}^* =& \sqrt{\frac{2}{F_n}}\cos\left(\frac{2\pi \mu j}{F_n}\right),
    \braket{j|\mu,-}^* = \sqrt{\frac{2}{F_n}}\sin\left(\frac{2\pi \mu j}{F_n}\right) \label{eq:pert_vec:S}
\end{align} 
By combining the gap labelling theorem and \eqref{eq:mcongruentmuApp}, we can compute the perpendicular space wavenumber of the slightly perturbed plane waves around each gap. $E_m$ is monotonic when $m \in [0,\lfloor F_n/2\rfloor]$. Therefore, for gaps with an even number of states below them
\begin{align}\label{eq:mtoqeven}
    &2m \equiv q F_{n-1} \equiv 2\mu F_{n-1} \pmod{F_{n}}\nonumber \\
    \implies& \mu =\frac{c F_n + q}{2}
\end{align}
where $c$ is some integer. For gaps with an odd number of states below them, an analogous calculation at the edge of the Brillouin zone \footnote{Using the convention that the eigenstate at the edge of the Brillouin zone with label $\mu$ has the wavenumber $\frac{2\pi (\mu + 1/2)}{F_n}$} results in:
\begin{align}\label{eq:mtoqodd}
    &2m+1 \equiv q F_{n-1} \equiv (2\mu+1) F_{n-1} \pmod{F_{n}}\nonumber\\
    \implies& \mu =\frac{c F_n + q-1}{2}
\end{align}

\section{Robustness of the charge oscillations}
\label{sec:Rob_app}
\begin{figure}
    \centering
    \includegraphics[width = \columnwidth]{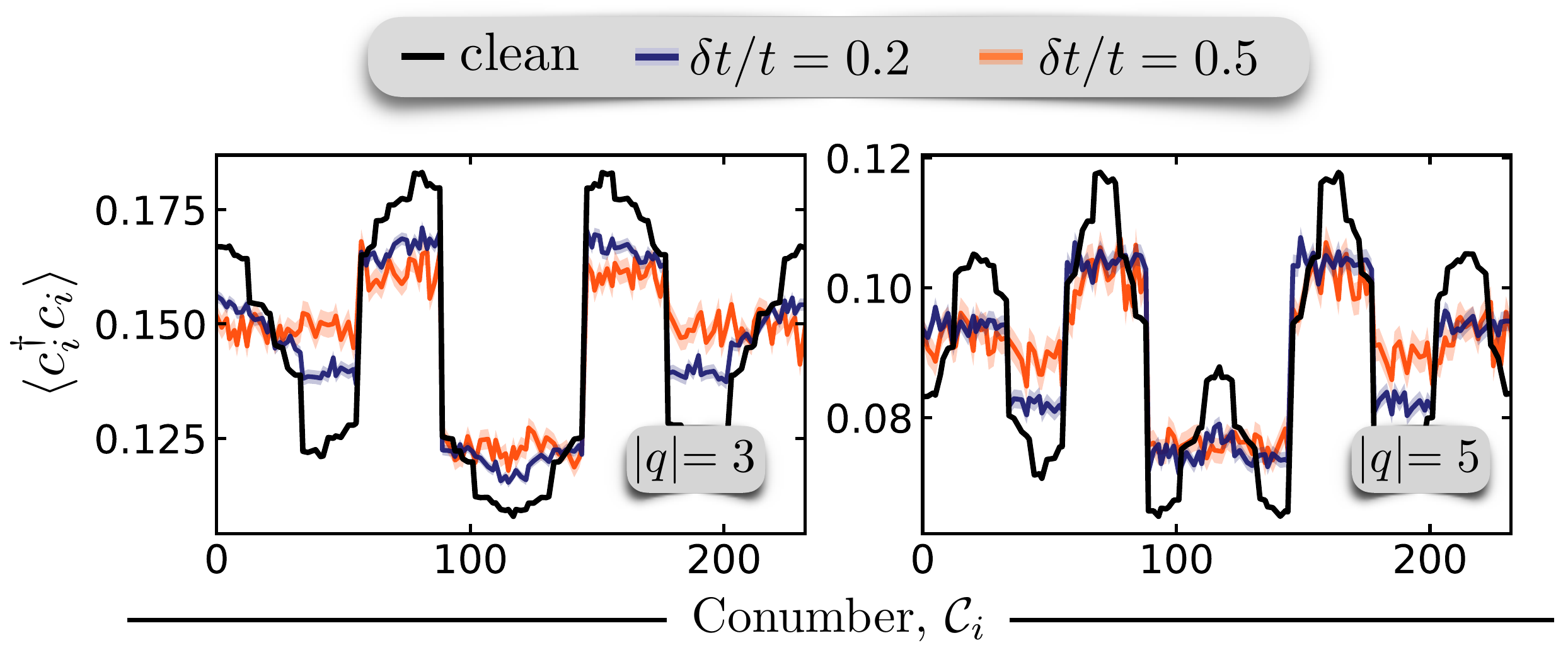} 
    \caption{Perpendicular space charge density for the disordered Fibonacci approximant at gaps with labels $|q|=3,5$ and varying disorder strength. Disorder averages are taken over 2000 realizations, the light background in the plots indicating the statistical error to the mean.}
    \label{fig:disorder_35}
\end{figure}

As mentioned in the main text, the oscillations of the charge density in perpendicular space are robust against local perturbations of the hopping integrals. The robustness of a particular oscillation at a gap $|q|$ is further found to scale inversely with the gap size $\Delta_{|q|}$, i.e., oscillatory features are lost for lower disorder strength in smaller gaps. In Fig.~\ref{fig:disorder_35}, we illustrate this qualitatively by the charge oscillations in the gaps $|q|=3$ and $|q|=5$, where $\Delta_{|q|=5} < \Delta_{|q|=3}$. For the gap with label $|q| = 3$, we find that the oscillations are fully visible upon introducing disorder of strength $\delta t/t = 0.2$, whereas oscillatory features are lost in the case $\delta t/t = 0.5$. For the $|q|=5$ gap, a disorder strength of $\delta t/t = 0.2$ is sufficient to wash out the five-period structure visible in the clean case, hence being less resilient to noise due to its smaller gap size.

\end{document}